\documentclass[a4paper,11pt]{article}
 
\usepackage{lineno}
\usepackage[colorlinks=false]{hyperref}
\usepackage{bm}
\usepackage{graphicx}
\usepackage{amssymb}
\usepackage{amsmath}
\usepackage{tabularx}
\usepackage{natbib}
\graphicspath{{./Figures/}}

\usepackage{caption}
\usepackage{amsmath}
\usepackage{amsfonts}
\usepackage{amssymb}
\newcommand{\cm}{{\mathcal M}}
\newcommand{\acc}[1]{\left\{#1\right\}}
\newcommand{\Esp}[1]{{\mathbb E}\left[ #1 \right]}
\newcommand{\ve}[1]{\bm{#1}}
\newcommand{\Ve}[1]{\bm{#1}}
\newcommand{\vx}{\ve{x}}
\newcommand{\vX}{\ve{X}}

\newcommand{\cx}{{\mathcal X}}
\newcommand{\cy}{{\mathcal Y}}
\newcommand{\ca}{{\mathcal A}}
\newcommand{\cn}{{\mathcal N}}
\newcommand{\cu}{{\mathcal U}}

\newcommand{\Nn}{{\mathbb N}}
\newcommand{\Rr}{{\mathbb R}}

\newcommand{\PsiAlpha}{\Psi_{\ve{\alpha}}}
\newcommand{\yAlpha}{y_{\ve{\alpha}}}

\newcommand{\tr}{^{\textsf T}}				

\DeclareMathOperator*{\argmin}{argmin}

\modulolinenumbers[5]

\usepackage[left=25mm,right=25mm,top=1.5cm,bottom=1.5cm,includeheadfoot]{geometry}
\title{An active-learning algorithm that combines sparse polynomial chaos 
	expansions and bootstrap for structural reliability analysis}

\author{S. Marelli and B. Sudret}

\begin{document}

\maketitle

\abstract{
Polynomial chaos expansions (PCE) have seen widespread use in the context of 
uncertainty quantification. However, their application to structural 
reliability problems has been hindered by the limited performance of PCE in the 
tails of the model response and due to the lack of local metamodel error 
estimates.
We propose a new method to provide local metamodel error estimates based on 
bootstrap resampling and sparse PCE.
An initial experimental design is iteratively updated based on the current 
estimation of the limit-state surface in an active learning algorithm. The 
greedy algorithm uses the bootstrap-based local error estimates for the 
polynomial chaos predictor to identify the best candidate set of points to 
enrich the experimental design. 
We demonstrate the effectiveness of this approach on a well-known analytical 
benchmark representing a series system, on a truss structure and on a complex 
realistic frame structure problem.\\

\noindent
\begin{tabular}{ll}
\textbf{Keywords:} &
Polynomial Chaos Expansions, Adaptive Designs, 
Bootstrap, Structural Reliability, \\
&Active Learning
\end{tabular}

}

\section{Introduction}
\label{sec:Introduction}
Structural reliability analysis aims at computing the probability of failure of 
a system with respect to some performance criterion in the presence of 
uncertainty in its structural and operating parameters.
Such uncertainty can be modelled by a random vector $\vX \in \Rr^M$ with 
prescribed joint probability density function $f_{\vX}$. 
The limit-state function $g$ is defined over the support of $\vX$ such that 
$\acc{\vx: g(\vx)\leq 0}$ defines the failure domain, while $\acc{\vx: g(\vx)> 
	0}$ defines the safe domain. The \textit{limit state surface} implicitly 
defined by $g(\vx) = 0$ lies at the boundary between the two domains.
The probability of failure of such a system can be defined as 
\citep{Melchers1999,Lemaire2009}:
\begin{equation}
\label{eqn:Pf_int}
P_F = \int_{\acc{\vx: g(\vx)\leq 0}} f_{\vX}(\vx) d\vx.
\end{equation}

A straightforward approach to compute the integral in Eq.~\eqref{eqn:Pf_int} is 
to use of Monte Carlo Simulation (MCS).
However, standard MCS approaches can often not be used in the presence of 
complex and computationally expensive engineering models, because of
the large number of samples they require to estimate small probabilities 
(typically in the order of $\sim 10^{k+2}$ for $P_F \approx 10^{-k}$) with 
acceptable accuracy. 
Well-known methods based on local approximation of the limit-state function 
close to the failure domain (such as FORM \citep{Hasofer1974} and SORM 
\citep{Rackwitz78}) can be more efficient, yet they are usually based on 
linearisation and tend to fail in real-case scenarios with highly non-linear 
structural models.

In contrast, methods based on surrogate modelling have gradually gained 
momentum in the last few years. Due to the nature of the problem of estimating 
low probabilities, most recent methods combine active-learning-based greedy 
algorithms with Gaussian process surrogate models (Kriging). Among the first 
works to propose this approach, the earliest applications in this context were 
the efficient global reliability 
analysis method (EGRA) by \citet{Bichon2008,Bichon2011}, and the 
active-learning reliability (AK-MCS) method based on Kriging by \citet{Echard2011}. 
More recently, Kriging has been employed to devise quasi-optimal importance 
densities in \citet{Dubourg2013,Dubourg2014}. Amongst other variations, 
polynomial-chaos-based Kriging has also been used as an alternative 
metamodelling technique \citep{SchoebiASCE2016} to overcome some of the 
limitations of pure Kriging-based methods.
Additional works on the topic of Kriging and structural reliability can be 
found, including extensions of the original AK-MCS algorithm to more advanced 
sampling techniques \citep{Echard2013,Balesdent2013}, system reliability 
\citep{Fauriat2014} and for the exploration of multiple-failure regions 
\citep{CadiniZio2014}.

Polynomial chaos expansions (PCE) \citep{Ghanembook1991} are a well-established 
tool in the context of uncertainty quantification, with applications in 
uncertainty propagation \citep{Xiu2002}, sensitivity analysis 
\citep{SudretHandbookUQ} and, to a lesser degree, structural reliability 
\citep{Sudret2002}. While often considered as an efficient surrogate modelling 
technique due to their global convergence behaviour, PCEs have been employed  
only seldom in reliability analysis (see, \textit{e.g.} \citet{Notin2010}) due 
to their lack of accuracy in the tails of the model response distribution, 
which are essential in this field. 

In addition, most active-learning approaches with surrogates require some form 
of local error estimate to adaptively enrich a small set of model 
evaluations close to the limit state surface. Kriging-based methods can rely 
on the Kriging variance for this task, but PCEs do not provide a natural 
equivalent.

In this paper, we leverage on the properties of regression-based sparse-PCE 
\citep{BlatmanJCP2011} to derive a local error estimator based on 
\textit{bootstrap resampling}. We then use this estimator to construct an 
active-learning strategy that adaptively approximates the limit-state function 
with PCE by minimizing a \textit{misclassification probability}-based learning 
function at every iteration. 
The method is then showcased on a standard benchmark functions representing a 
series system and on a realistic structural frame engineering example.

\section{Methodology}
\label{sec:Methodology}
\subsection{Polynomial Chaos Expansions}
\label{sec:Methodology:PCE}

Consider a finite variance model $Y = \cm(\vX)$ representing the response of 
some quantity of interest (QoI) $Y$ to the random input parameters $\vX \in 
\Rr^M$, modelled by a joint probability distribution function (PDF) $f_{\vX}$. 
Also consider the functional inner product defined by:
\begin{equation}\label{eqn:inner product}
	\left<g,h\right> \equiv \int_{\vx \in\Omega_{\vX}} 
	g(\vx)h(\vx)f_{\vX}(\vx)d\vx 
	= \Esp{g(\vX) h(\vX)}
\end{equation}
where $\Omega_{\vX}$ represents the input domain. Under the assumption of 
independence of the input variables, that is $f_{\vX}(\vx) = 
\prod\limits_{i=1}^M f_{X_i}(x_i)$, one can represent $\cm(\vX)$ as the 
following \textit{generalised polynomial chaos expansion}  (see, \textit{e.g.} 
\citet{Ghanembook1991,Xiu2002}):
\begin{equation}
\label{eqn:PCE}
Y = \cm(\vX) = \sum\limits_{\ve{\alpha}\in \Nn^M} 
y_{\ve{\alpha}}\Psi_{\ve{\alpha}}(\vX),
\end{equation}
where the $\yAlpha$ are real coefficients and $\ve{\alpha}$ is a multi-index 
that identifies the degree of the multivariate polynomial $\PsiAlpha$ in each 
of the input variables $X_i$:
\begin{equation}
\label{eqn:PsiAlpha}
\PsiAlpha = \prod\limits_{i = 1}^M \phi^{(i)}_{\alpha_i}(X_i).
\end{equation}
Here $\phi^{(i)}_{\alpha_i}$ is a polynomial of degree $\alpha_i$ 
that belongs to the family of orthogonal polynomials w.r.t. the marginal PDF 
$f_{X_i}$. 
For more details on the construction of such polynomials for both standard and 
arbitrary distributions, the reader is referred to \citet{Xiu2002}. 

In the presence of a complex dependence structure between the input 
variables, it is always possible to construct isoprobabilistic transforms 
(\textit{e.g.} Rosenblatt or Nataf transforms, see \textit{e.g.} 
\citet{Lebrun2009c}) to decorrelate the input variables prior to the expansion, 
even in the case of complex dependence modelled by vine 
copulas \citep{Torre2017_arxiv}.  
For the sake of notational simplicity and without loss of generality, we will 
hereafter assume independent input variables. 

In practical applications, the series expansion in Eq.~\eqref{eqn:PCE} is 
traditionally truncated based on the maximal degree $p$ of the expansion, thus 
yielding a set of basis elements identified by the multi-indices ${\ve{\alpha} 
\in \ca: \sum\limits_{i = 1}^M \alpha_i\leq p}$, with $\text{card}(\ca) \equiv 
P = \binom{M+p}{p}$, or using more advanced truncation schemes that favour 
sparsity, \textit{e.g.} hyperbolic truncation \citep{BlatmanPEM2010}. 
The corresponding expansion coefficients $\ve{\yAlpha}$ can then be calculated 
efficiently via least-square analysis based on an existing sample of the input 
random vector $\cx = \acc{\vx^{(1)},\cdots,\vx^{(N)}}$, known as the 
\textit{experimental design} (ED), and the corresponding model responses $\cy = 
\acc{y^{(1)},\cdots,y^{(N)}}$ as follows:
\begin{equation}
\label{eqn:LSQ PCE}
\ve{\yAlpha} = \text{argmin}\frac{1}{N} \sum\limits_{i = 1}^N
\left[
y^{(i)} - \sum\limits_{\ve{\alpha}\in \ca} \yAlpha\PsiAlpha(\vx^{(i)})
\right]^2.
\end{equation}

When the number of unknown coefficients $P$ is high (\textit{e.g.} for 
high-dimensional inputs or high-degree expansions), regression strategies that 
favour sparsity are needed to avoid over-fitting in the presence of a 
limited-size experimental design and to make the analysis at all feasible with 
a reasonable sample size $N$. 
Amongst them, \textit{least angle regression} (LARS, \citet{Efron2004}), based 
on a regularized version of {Eq. \eqref{eqn:LSQ PCE}}, has proven to be 
very effective in tackling realistic engineering problems even in relatively 
high dimensions (\textit{i.e.} $M \sim 100$). In this paper, we adopt the 
full degree-adaptive, sparse PCE based on hybrid-LARS introduced in 
\citet{BlatmanJCP2011}), as implemented in the \textsc{UQLab} Matlab software 
(\citep{Marelli2014,UQdoc_10_104}).

\subsection{Bootstrap-based local error estimation in PCE}
\label{sec:Methodology:Local error}
\subsubsection{Bootstrap in least-square regression}
Adopting a least-square regression strategy to calculate the coefficients in 
Eq.~\eqref{eqn:LSQ PCE} allows one to use the \textit{bootstrap} resampling 
method \citep{Efron1982} to obtain information on the variability 
in the estimated coefficients due to the finite size of the experimental 
design. 
Suppose that a set of estimators $\ve{\theta}$ is a function of a finite-size 
sample $\cx = \acc{\vx^{(1)},\cdots,\vx^{(N)}}$ drawn from the random vector 
$\vX$. Then the \textit{bootstrap} method consists in drawing $B$ new sample 
sets $\acc{\cx^{(1)},\cdots,\cx^{(B)}}$ from the original $\cx$ by 
\textit{resampling with substitution}. This is achieved by randomly assembling 
$B-$times $N$ realizations $\vx^{(i)}\in \cx$, possibly including repeatedly 
the same realization multiple times within each sample. The set of estimated 
quantities can then be re-calculated from each of the $B$ samples, thus 
yielding a set of estimators $\Ve{\Theta} = 
\acc{\ve{\theta}^{(1)},\cdots,\theta^{(B)}}$. 
This set of estimators can then be used to directly assess the variability of 
$\Ve{\theta}$ due to the finite size of the experimental design $\cx$, at no 
additional costs, \textit{e.g.} by calculating statistics, or directly using 
each realization separately. 
Application of the bootstrap method combined with PCE to provide confidence 
bounds in the estimated $P_F$ in structural reliability applications can be 
found  in \textit{e.g.} \citet{Notin2010,Picheny2010a}.

\subsubsection{Bootstrap-PCE}
\label{sec:Methodology:bPCE}
We propose to use the bootstrap technique to provide local error estimates to 
the PCE predictions. The rationale is the following: the PCE coefficients
$\ve{\yAlpha}$ in Eq.~\eqref{eqn:LSQ PCE} are estimated from the experimental 
design $\cx$, therefore they can be resampled through bootstrap. 
This can be achieved by first generating a set of bootstrap-resampled 
experimental designs $\acc{\cx^{(b)},\cy^{(b)}, b = 1,\cdots,B}$. 
For each of the generated designs, one can calculate a corresponding set of 
coefficients $\ve{\yAlpha}^{(b)}$, effectively resulting in a set of $B$ 
different PCEs. 
Correspondingly, the response of each PCE can be evaluated at a point $\vx$ as 
follows:
\begin{equation}
\label{eqn:bPCE}
Y_{PC}^{(b)}(\vx) = \sum\limits_{\ve{\alpha}\in \ca} 
y^{(b)}_{\ve{\alpha}}\,\Psi_{\ve{\alpha}}(\vx),
\end{equation}
thus yielding a full response sample at each point $\acc{Y_{PC}^{(b)}(\vx), b = 
1,\cdots,B}$. 
Therefore, empirical quantiles can be employed to provide local error bounds on 
the PCE prediction at each point, as well as to any derived quantity 
(\textit{e.g.} $P_F$ or sensitivity indices, see \textit{e.g.} 
\citet{Picheny2010a,Dubreuil2014}).

This bootstrap-resampling strategy in Eq.~\eqref{eqn:bPCE} yields in 
fact a family of $B$ surrogate models that can be interpreted as 
\textit{trajectories}. Figure~\ref{fig:bPCE resampling} showcases how such 
trajectories can be directly employed to assess confidence bounds on point-wise 
predictions on a simple 1D test function given by:
\begin{equation}\label{eqn:1D Sinc}
f(x) = x\sin(x),\qquad x\in [0, 2\pi],
\end{equation}
where the single random variable is assumed to be uniformly distributed within 
the bounds $X\sim\cu(0,2\pi) $, and where $B = 100$ bootstrap samples have been 
used.

\begin{figure}
	\begin{center}
		\includegraphics[width=.65\textwidth]{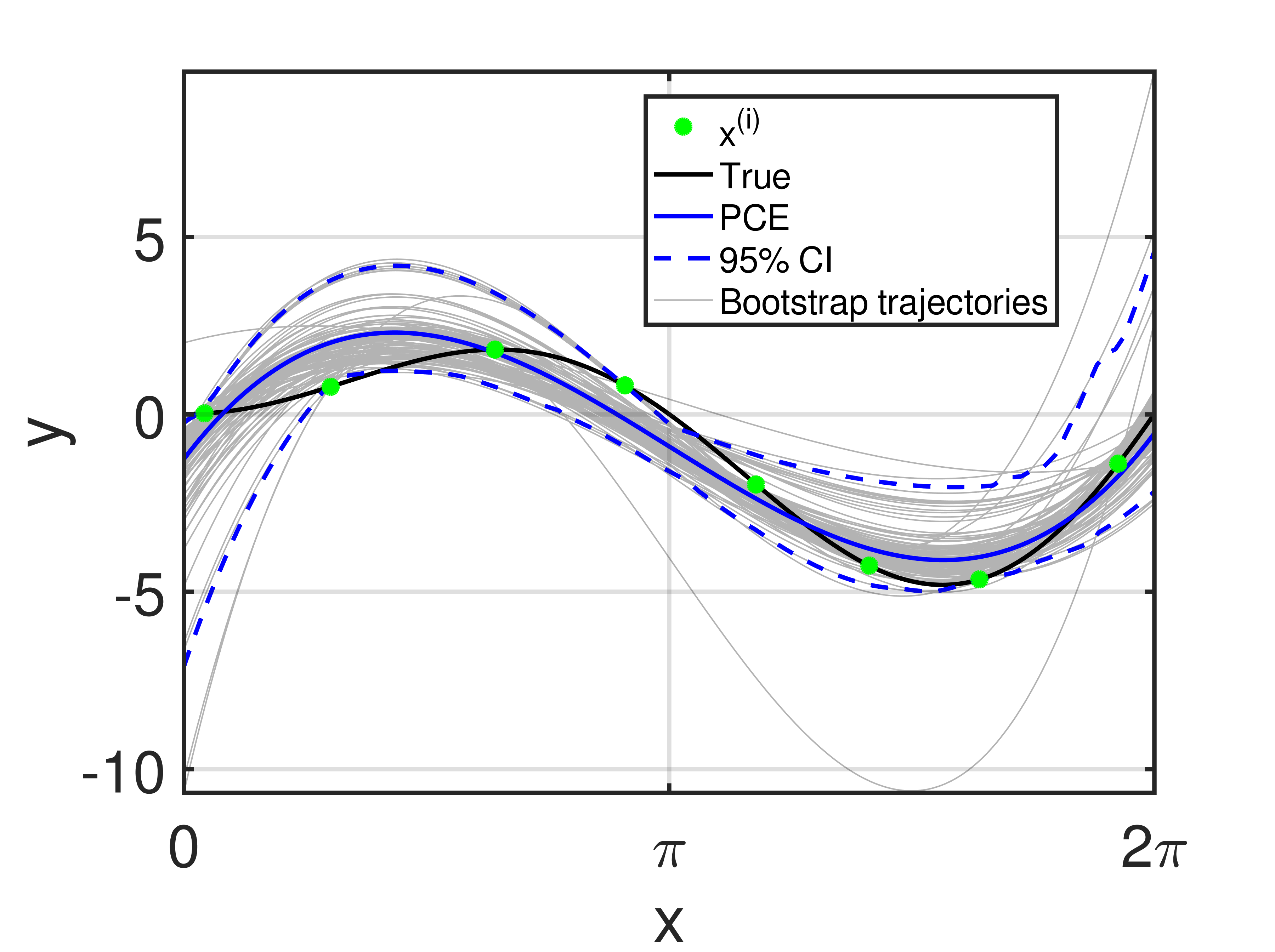}
	\end{center} 
\caption{Bootstrap-resampled trajectories ($B = 100$) of the simple 1D 
analytical function in Eq.~\eqref{eqn:1D Sinc}. 
The black line represents the true model, sampled at the 8 experimental design 
points $\vx^{(i)}$ (green dots).
The PCE surrogate is represented by the blue line, while the bootstrap 
trajectories are given by the gray lines. The corresponding 95\% empirical 
inter quantile-range is given by the dashed blue lines.}\label{fig:bPCE 
resampling}
\end{figure}

This process of bootstrap-based trajectory resampling to provide better 
estimates of point-wise confidence bounds has been recently explored in the 
Gaussian process modelling literature, see \textit{e.g.}, 
\citet{Hertog2006,Beers2008}.

We refer to this approach as to \textit{bootstrap-PCE}, or bPCE in short.

\subsubsection{Fast bPCE}
\label{sec:Methodology:fast bPCE}
Because the training of a PCE model with sparse least-square analysis may be 
time consuming, especially in high dimension and/or when an already large 
experimental design is available (\textit{i.e.} $N\sim 10^3$), and because in 
this particular application we do not need very accurate estimates on the 
bounds of the derived quantities, we adopt a \textit{fast bPCE} approach.
In this approach, the sparse polynomial basis identified by the LARS algorithm 
during calibration is calculated only once from the available full experimental 
design $\cx$, and bootstrapping is applied only to the final \textit{hybrid} 
step, which consists in a classic ordinary least-square regression on the 
sparse basis \citep{BlatmanJCP2011}.

In the presence of a very expensive model, however (\textit{i.e.} requiring 
several hours for a single model run), we recommended to adopt full 
bootstrapping, including the estimation of the sparse PCE basis for each 
of the $B$ bootstrapped experimental designs $\cx^{(1,\cdots,B)}$.

\subsection{Active bPCE-based reliability analysis}
\label{sec:Methodology:bPCE reliability}
In this section we present an adaptation of the Adaptive PC-Kriging MCS 
algorithm in \citet{SchoebiASCE2016} (based in turn on the original AK-MCS 
algorithm by \citet{Echard2011}), that makes use of the bPCE just introduced. 
Consistently with \citet{Echard2011,SchoebiASCE2016}, in the following we will 
refer to this algorithm as \textit{active bootstrap-polynomial-chaos 
Monte-Carlo simulation} (A-bPCE).
We follow the original idea of adaptively building a surrogate of the 
limit-state function starting from a small initial experimental design and 
subsequently refining it to optimize the surrogate performance for structural 
reliability. The ultimate goal of the adaptation is to retrieve an estimate of 
$P_F$ that is comparable to that of a direct Monte Carlo simulation (MCS) using 
a large sample set with a much smaller experimental design. 
The algorithm is summarized as follows:
\begin{enumerate}
	\setcounter{enumi}{-1}
	\item \textit{Initialization}:
	\begin{enumerate}
		\item Generate an initial experimental design (\textit{e.g.} through 
		Latin hypercube sampling or uniform sampling of a ball 
		\citep{DubourgThesis}) and calculate the corresponding bPCE surrogate 
		(see Section~\ref{sec:Methodology:Initial design}).
		\item Generate a large reference MCS sample $\cx_{MCS} = 
		\acc{\vx_{MCS}^{(i)},~i = 1,\cdots,N_{MCS}}$ of size $N_{MCS}$ 
		(\textit{e.g.} $N_{MCS} = 10^6\gg N$). A discussion on the choice of a 
		suitable MCS sample is given in	Section~\ref{sec:Methodology:Inner 
		MCS}).
	\end{enumerate}
	\item\label{itm:update} Calculate a set of MCS estimators of the 
	probability of 
	failure: $\acc{\widehat{P}_F^{(b)},~b=1,\cdots,B}$ with the current 
	bPCE surrogate. 
	\item Evaluate one or more suitable convergence criteria on 
	$\widehat{P}_F^{(1,\cdots,B)}$ 	(see 	
	Section~\ref{sec:Methodology:Convergence criteria}). 
	If they are met, go to Step \ref{itm:termination} (terminate the 
	algorithm). Otherwise continue to the next step.
	\item Evaluate a suitable learning function on the MCS sample $\cx_{MCS}$ 
	(see Section~\ref{sec:Methodology:Learning function}). 
	Choose one or more additional $\vx_{MCS}^{(i)}\in \cx_{MCS}$ and add them 
	to the ED (see Section~\ref{sec:Methodology:ED enrichment}).
	\item Update the bPCE surrogate on the new ED and return to 
	Step~\ref{itm:update} 
	\item\label{itm:termination}\textit{Algorithm termination}: return the 
	$\widehat{P}_F$ resulting from the PCE on the current ED, as well as the 
	error bounds derived \textit{e.g.} from the extremes or the empirical 
	quantiles of the current $\widehat{P}_F^{(1,\cdots,B)}$ set. 
\end{enumerate}

A detailed description of each step of the algorithm is given in the following 
sections.

\subsubsection{Initial experimental design}
\label{sec:Methodology:Initial design}
The initial experimental design is usually generated by space-filling sampling 
techniques of the random vector $\vX$, such as Latin hypercube sampling (LHS) 
or pseudo-random sequences (\textit{e.g.} Sobol' sequence). Alternative 
sampling techniques, such as the uniform sampling of a ball, have also  
proven effective in the context of structural reliability when low 
probabilities of failure are expected \cite{DubourgThesis}. Note that this 
initial set of model evaluations does not need to be a subset of the reference 
sample $\cx_{MCS}$ used later to evaluate the $P_F$ estimates during the 
iterations of the algorithm.

\subsubsection{Inner MCS-based estimate of $P_F$}
\label{sec:Methodology:Inner MCS}
While the estimation of the $P_F$ via MCS is trivial, as it simply entails 
counting the number of samples that belong to the failure domain, some 
discussion about the number of samples $N_{MCS}$ in this step is needed. 
Throughout this paper, we opted to choose a single MCS sample $\cx_{MCS}$ large 
enough to ensure a relatively small CoV for the $P_F$ estimate at every 
iteration. This is by no means a requirement of this algorithm, but it 
simplifies significantly the notation (because $\cx_{MCS}$ becomes independent 
on the current iteration) and in some cases (as noted in both 
\citet{Echard2011} and \citet{SchoebiASCE2016}) it can result in stabler 
convergence, due to the lowered MCS noise in the estimation of $P_F$ during 
each iteration. 
This technique is known as \textit{common random numbers} in the context of 
repeated reliability analysis \textit{.e.g.} in reliability-based design 
optimization \citep{Taflanidis2008}.
It is entirely possible to redraw the $\cx_{MCS}$ during every 
iteration, possibly each time with a different number of samples $N_{MCS}$.

The choice of $N_{MCS} = 10^6$ ensures that the CoV estimated probabilities of 
failure in the order of $P_F\geq 10^{-3} $ is always smaller than $5\%$, which 
we found suitable in our application examples. The choice of a single MCS 
sample drawn during the algorithm initialization also allows us to 
use the application examples to focus more on the convergence of the active 
learning part of A-bPCE, which is the focus of this paper.

In more general applications, the order of magnitude of $P_F$ may be unknown. 
In this case, it is recommended instead to set a target desired CoV for the 
estimation of $P_F$ at each iteration (as proposed in the original AK-MCS 
algorithm in \citet{Echard2011}), and gradually add samples to $\cx_{MCS}$ until 
it is reached. 

\subsubsection{Convergence criteria}
\label{sec:Methodology:Convergence criteria}
The proposed convergence criterion of choice is directly inspired by 
\citet{SchoebiASCE2016,Notin2010} and it depends on the stability of the $P_F$ 
estimate at the current iteration. 
Let us define:
\begin{equation}
\label{eqn:Pf+-}
\begin{split}
\widehat{P}_F^+ &
= \max\limits_{b = 1,\cdots,B}\left(\widehat{P}_F^{(b)}\right)\\
\widehat{P}_F^- &
= \min\limits_{b = 1,\cdots,B}\left(\widehat{P}_F^{(b)}\right).
\end{split}
\end{equation}
Convergence is achieved when the following condition is satisfied for at least 
two consecutive iterations of the algorithm:
\begin{equation}
\label{eqn:Pf convergence}
\frac{\widehat{P}_F^+ - \widehat{P}_F^-}{\widehat{P}_F} \leq 
\epsilon_{\widehat{P}_F},
\end{equation}
with $0.05\leq\epsilon_{\widehat{P}_F}\leq 0.15$ in typical usage scenarios. 

\subsubsection{Learning function}
\label{sec:Methodology:Learning function}
A learning function is a function that allows one to rank a set of candidate 
points based on some utility criterion that depends on the desired application. 
In this case, we adopt the same heuristic approach proposed in 
\citet{SchoebiASCE2016}, by focusing on the probability of misclassification of 
the bPCE model on the candidate set given by $\cx_{MCS}$. 

Due to the availability of the bootstrap response samples 
$\cy_{MCS}^{(1,\cdots,B)}$, it is straightforward to define a measure of the 
misclassification probability $U_{FBR}$ (where the subscript FBR stands for 
\textit{failed bootstrap replicates}) at each point $\vx_{MCS}^{(i)}\in 
\cx_{MCS}$ as follows:
\begin{equation}
\label{eqn:U function}
U_{FBR}(\vx_{MCS}^{(i)}) = 
\left|\frac{B_S(\vx_{MCS}^{(i)})-B_F(\vx_{MCS}^{(i)})}{B}\right|
\end{equation}
where $B_S(\vx_{MCS}^{(i)})$ and $B_F(\vx_{MCS}^{(i)})$ are the number of safe 
(resp. failed) bPCE replicate predictions at point $\vx_{MCS}^{(i)}$ (with 
$B_S(\vx_{MCS}^{(i)}) + B_F(\vx_{MCS}^{(i)}) = B$). 
When all the $B$ replicates consistently classify $\vx_{MCS}^{(i)}$ in the safe 
or in the failure domain, $U_{FBR} = 1$ (minimum misclassification probability).
In contrast, $U_{FBR} = 0$ corresponds to the case when the replicates are 
equally distributed between the two domains. 
In the latter case, 50\% of the $B$ bootstrap PCEs predict that 
$\vx_{MCS}^{(i)}$ is in the safe domain, while the other 50\% predicts that 
$\vx_{MCS}^{(i)}$ belongs to the failure domain. 
Therefore, maximum epistemic uncertainty on the classification of a point 
$\ve{x}^{(i)}_{MCS}$ is attained when $U_{FBR}$ is minimum.

\subsubsection{Enrichment of the experimental design}
\label{sec:Methodology:ED enrichment}
The aim of the iterative algorithm described in 
Section~\ref{sec:Methodology:bPCE reliability} is to obtain a 
surrogate 
model that minimizes the misclassification probability. As a consequence, the 
learning function in Eq.~\eqref{eqn:U function} can be directly used to obtain 
a single-point enrichment criterion. The next best candidate point for the ED  
$\vx^*\in\cx_{MCS}$ is given by:
\begin{equation}
\label{eqn:single point enrichment}
\vx^* = \argmin\limits_{\vx^{(i)} \in 
\cx_{MCS}}\left(U_{FBR}(\vx^{(i)})\right).
\end{equation}

Due to the global character of regression-based PCE, it can be beneficial to 
add multiple points in each iteration to sample several interesting regions of 
the parameter space simultaneously. 
The criterion in Eq.~\eqref{eqn:single point enrichment} can be 
extended to include $K$ distinct points simultaneously by following the 
approach in \citet{SchoebiASCE2016}.
A \textit{limit state margin} region is first defined as the set of points 
such that $U_{FBR}<1$ (\textit{i.e.} those point with non-zero 
misclassification probability at the current iteration). 
Subsequently, $k$-means clustering techniques (see, \textit{e.g.}, 
\citet{Zaki2014}) can be used at each iteration to identify $K$ disjoint 
regions $\acc{\cx^{(1,\cdots,K)}_{MCS}}$ in the limit-state margin. Then, 
Eq.~\eqref{eqn:single point enrichment} can be directly applied to each of the 
subregions to obtain $K$ different enrichment points:
\begin{equation}
\label{eqn:multi-point enrichment}
\vx^{*}_k = \argmin\limits_{\vx^{(i)} \in 
	\cx_{MCS}^{(k)}}\left(U^{(k)}_{FBR}(\vx_k^{(i)})\right), \quad k = 
1,\cdots,K
\end{equation}
where $\vx^{*}_k\in \cx^{(k)}_{MCS}$ is the $k-$th enrichment sample and 
$U^{(k)}_{FBR}$ is the learning function evaluated on the $k-$th region of the 
parameter space.

Note that this approach is also convenient when parallel computing facilities 
are available and in the presence of computationally expensive objective 
functions, as the evaluation of the $K$ enrichment points can be carried out 
simultaneously.

\section{Results on benchmark applications}
\label{sec:Results}

All the algorithm development and the final calculations presented in this 
section were performed with the polynomial chaos expansions and reliability 
analysis modules of the  \textsc{UQLab} software for uncertainty quantification 
\citep{Marelli2014,UQdoc_10_104,UQdoc_10_107}.

\subsection{Series system}
\label{sec:Results:Four branch}
A common benchmark for reliability analysis functions is given by the 
four-branch function, originally proposed in \citet{Waarts2000}, that represents 
a series system comprising four components with different failure criteria.
Although it is a simple analytical function, it shows multiple failure regions 
and a composite limit-state surface. Its two-dimensional limit state function 
reads:
\begin{equation}
\label{eqn:four branch}
g(\vx) = \min\left\{ 
\begin{matrix}
3+0.1(x_1+x_2)^2 - \frac{x_1 + x_2}{\sqrt{2}}\\
3+0.1(x_1+x_2)^2 + \frac{x_1 + x_2}{\sqrt{2}}\\
(x_1-x_2) + \frac{6}{\sqrt{2}}\\
(x_2-x_1) + \frac{6}{\sqrt{2}}\\
\end{matrix}	
\right\}
\end{equation}
where the two random input variables $X_1\sim \cn(0,1)$ and $X_2\sim\cn(0,1)$ 
are modelled as independent standard normals. Failure occurs when $g(\vx)\leq 
0$. 

\begin{figure}
	\begin{minipage}{.45\textwidth}
		\centering\small
		\includegraphics
		[width = \textwidth,clip=true,trim=60 0 90 20]
		{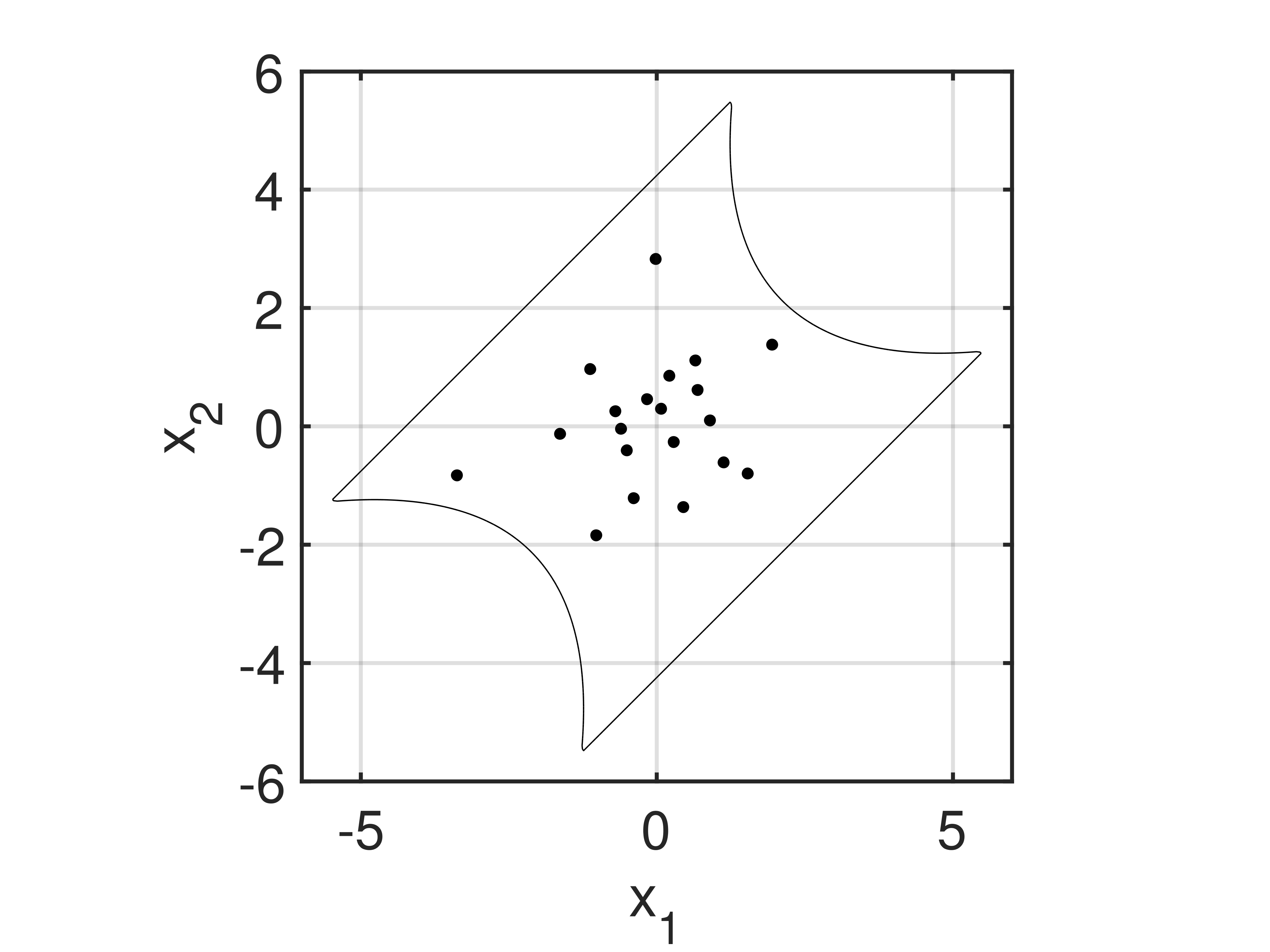}\\
		a. \textit{Initial experimental design}
	\end{minipage}
	\begin{minipage}{.45\textwidth}
		\centering\small
		\includegraphics
		[width = \textwidth, clip=true,trim=60 0 90 20]
		{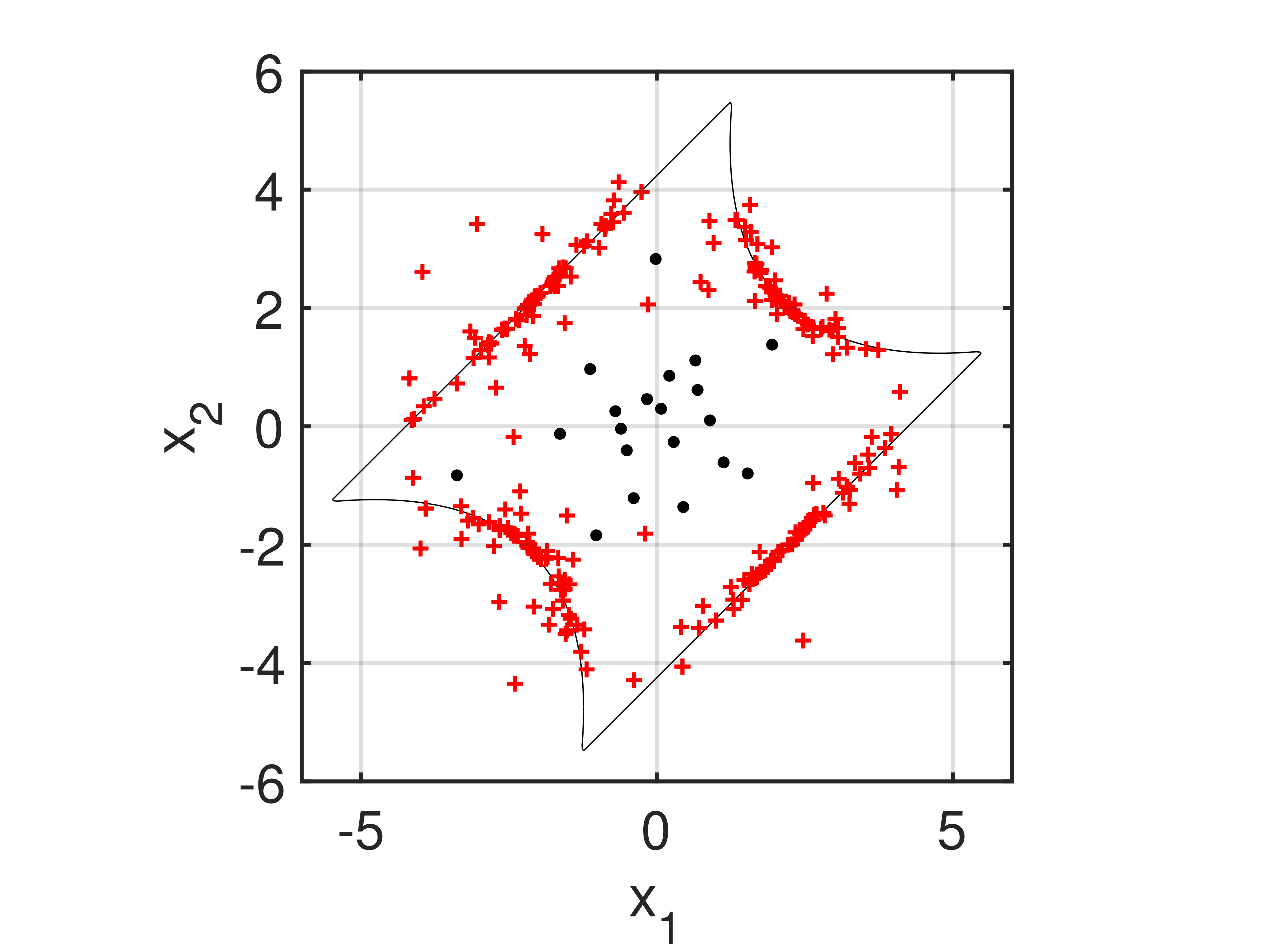}\\
		b. \textit{Final experimental design }
	\end{minipage}
	\caption{Four branch function: limit state surface (black line) and 
		experimental design before (black circles) and after (red crosses) 
		enrichment}
	\label{fig:Four branch 2D}
\end{figure}

Due to the multi-failure shape of the limit-state surface (represented as a 
solid black line in Figure~\ref{fig:Four branch 2D}), classic methods like 
FORM/SORM and importance sampling tend to fail with this benchmark problem. The 
reference failure probability of $P_F = 4.460\cdot 10^{-3}$ is obtained 
through an extremely large MCS ($N_{MCS} = 10^8$).

The initial experimental design for the A-bPCE algorithm was obtained with a 
space-filling LHS sample consisting of $N_{ini} = 20$ points drawn from the 
input distributions (black dots in Figure~\ref{fig:Four branch 2D}). Three 
points at a time were added to the experimental design during the enrichment 
phase of the algorithm. The number of replications for the A-bPCE algorithm 
is set to $B = 100$. After extensive testing, the algorithm was found to 
be very weakly dependent on the number of bootstrap replications, provided a 
minimum of $B \geq 20$ was provided. Indeed, the boostrap samples are used to 
identify areas of relatively large prediction variability, but an accurate 
estimate of such variability is never really needed by the algorithm. 
Degree adaptive sparse PCE (with maximum degree in the range $p\in[2, 10]$) 
based on LARS \citep{BlatmanJCP2011} was used to calibrate the PCE metamodel 
at each iteration. 
For validation and comparison purposes, a similar analysis was 
performed on the same initial ED with the AK-MCS module of \textsc{UQLab}, with 
an anisotropic Mat\'ern 5/2 ellipsoidal multivariate 
Kriging correlation function 
\citep{Rasmussen2006,UQdoc_10_107,UQdoc_10_105}. 
The convergence criterion in Eq.~\eqref{eqn:Pf convergence} was set to 
$\epsilon_{\widehat{P}_F} = 0.05$ for both the AK-MCS and A-bPCE algorithms.

\begin{figure}
	\centering
		\centering
		\includegraphics
		[width=.45\textwidth,clip=true,trim=0 0 22 00]
		{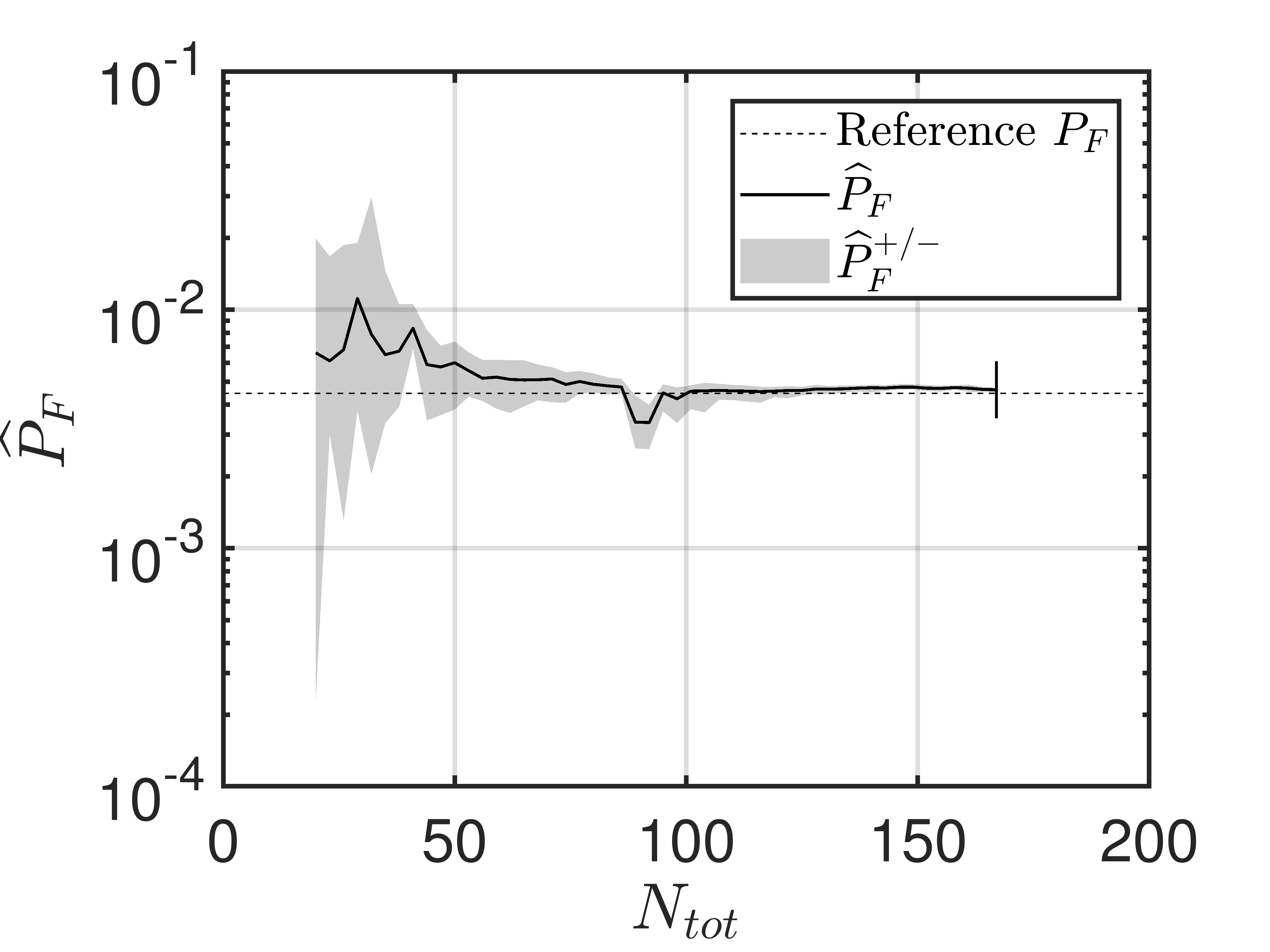}
		\includegraphics
		[width=.45\textwidth,clip=true,trim=0 0 22 00]
		{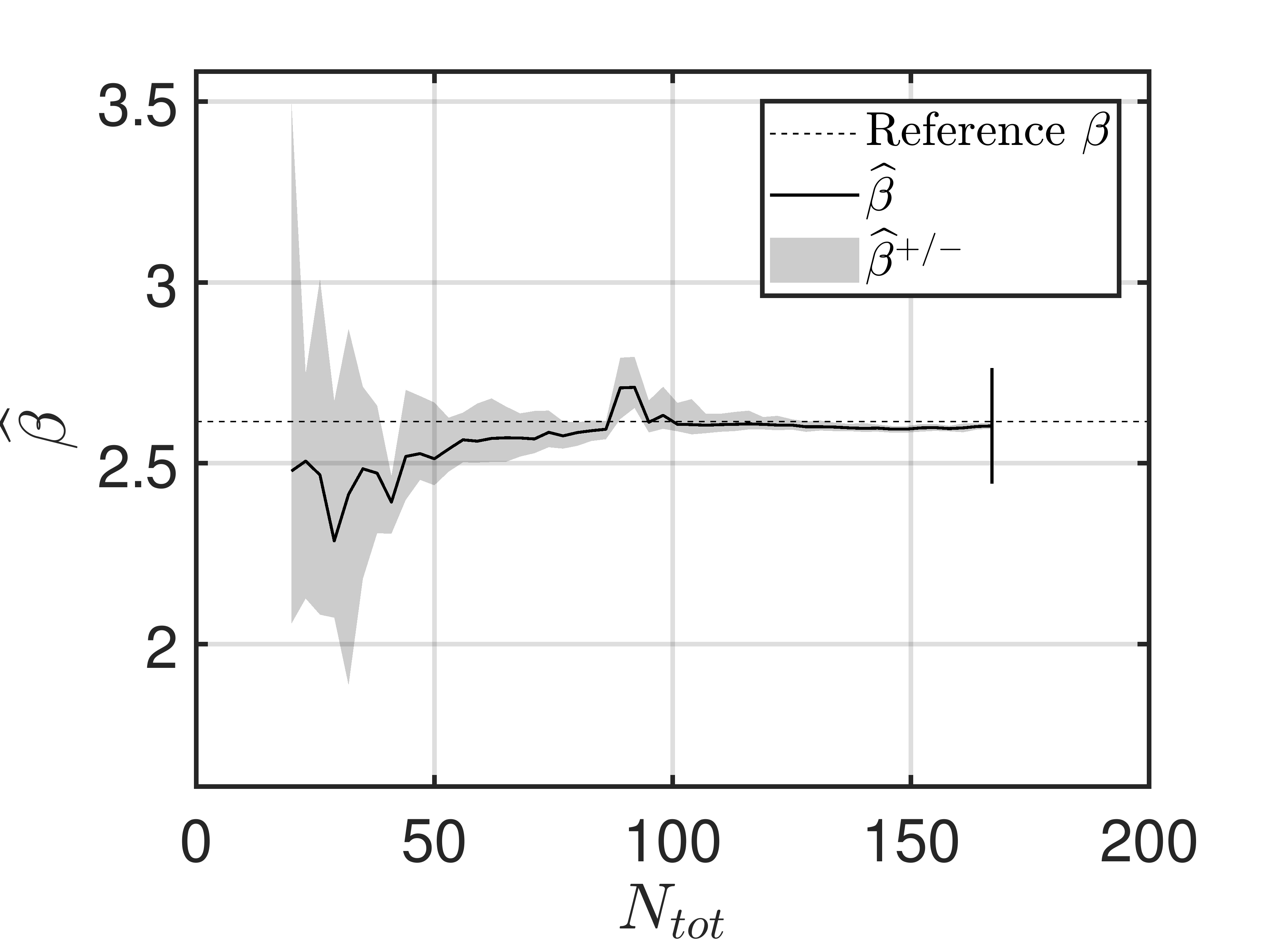}
                \caption{Convergence curves of the four-branch
                  limit-state function. The reference $P_F$ and $\beta$
                  are given by dotted lines.}
	\label{fig:Results:Four branch convergence}
\end{figure}

Convergence was achieved after 49 iterations, resulting in a total cost 
(including the initial experimental design) of $N_{tot} = 185$ model 
evaluations. 
The experimental design points added during the iterations are marked by red 
crosses on panel (b) of Figure~\ref{fig:Four branch 2D}. As expected, the 
adaptive algorithm tends to enrich the experimental design close to the limit 
state surface as it is adaptively learned during the iterations. 
A graphical representation of the convergence of the algorithm is shown in 
Figure~\ref{fig:Results:Four branch convergence}, where the estimated 
$\widehat{P}_F$ is plotted against the total number of model evaluations 
$N_{tot}$. The shaded area represents the $95\%$ confidence bounds based on the 
empirical quantiles as estimated from the bootstrap sample.

The final results of the analysis are summarized in Table~\ref{tbl:Results:Four 
branch}, where the generalised reliability index $\beta = -\Phi^{-1}(P_F)$ 
is also given for reference. 
For comparison, the reference MCS probability as well as an estimate from 
AK-MCS are also given. The latter converged to a comparably accurate 
estimate of $P_F$, at the cost of a slightly higher number of model evaluations.
Note that for engineering purposes, the algorithm could have been stopped 
earlier, \textit{i.e.} when a $5\%$ accuracy on the generalized reliability 
index is attained. In this case, the algorithm would have converged to 
a comparable result ($\widehat{\beta} = 2.56$) with only 50 runs of the model.
The final sparse PCE model after enrichment contained a total of $P = 12$ basis 
elements of degree up to $p = 5$.

\begin{table}
	\centering\small
	\caption{Comparison of different reliability analysis methods for the four 
		branch function}
	\begin{tabular}{cccccc}
		\hline
		Algorithm &
		$\widehat{P}_F$ &
		$[\widehat{P}_F^-, \widehat{P}_F^+]$&
		$\widehat{\beta}$&
		$[\widehat{\beta}^-, \widehat{\beta}^+]$&
		$N_{\text{tot}}$\\\hline
		\vspace{-.4cm}\\
		MCS (ref.) &
		$4.46\times 10^{-3}$&
		-&
		$2.62$&
		-&
		$10^8$\\
		AK-MCS &
		$4.52\times 10^{-3}$&
		$[4.38, 4.65]\times 10^{-3}$&
		$2.61$&
		$[2.60, 2.62]$&
		$200$\\
		
		A-bPCE &
		$4.62\times 10^{-3}$&
		$[4.5, 4.7]\times 10^{-3}$&
		$2.60$&
		$[2.59, 2.61]$&
		$167$\\\hline
	\end{tabular}
	\label{tbl:Results:Four branch}
\end{table}

\subsection{Two-dimensional truss structure}
\label{sec:Results:Truss}
\begin{figure}\centering
	\includegraphics[width=.9\textwidth]{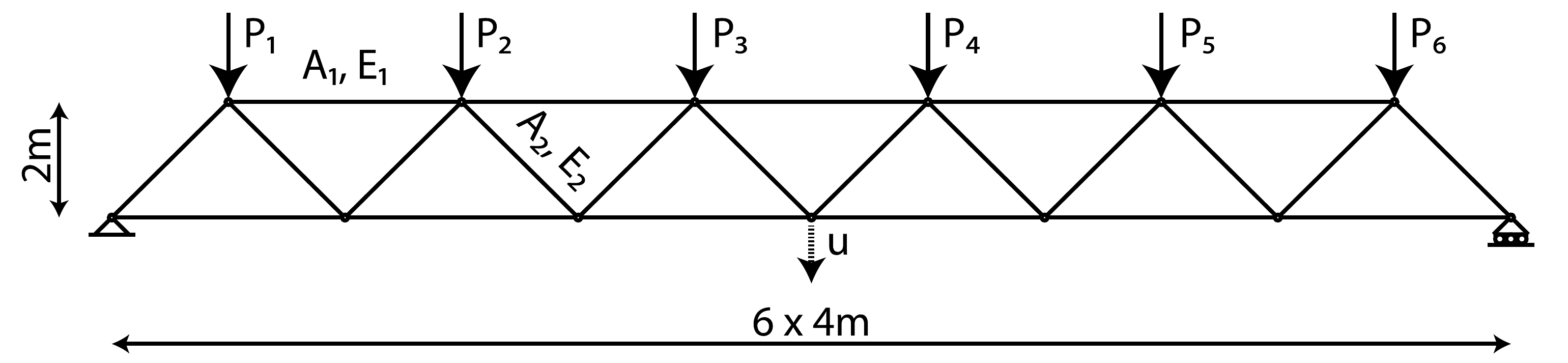}
	\caption{Two-dimensional truss structure with uncertain parameters. 
		Probability 
		distributions of the geometrical parameters $\acc{A_i,E_i}$ and of the 
		loads 
		$\acc{P_1,\cdots,P_6}$ are given in Table~\ref{tbl:Truss 
			input}}\label{fig:Truss}
\end{figure}

To test the algorithm on a more realistic engineering benchmark, consider the 
two-dimensional truss structure sketched in Figure~\ref{fig:Truss}.
This structure has been previously analysed in several works, see
\textit{e.g.} \citep{BlatmanJCP2011,BlatmanRESS2010,SchoebiASCE2016}. The truss 
comprises 23 bars and 13 nodes, with deterministic geometry yet with uncertain
material properties and random loads. 
The components of the input random vector $\vX = 
\left[A_{1,2},E_{1,2},P_{1,...,6}\right]\tr$ include the cross-section and the 
Young's modulus $\acc{A_1, E_1}$ of the horizontal bars, the cross-section and 
the Young's modulus $\acc{A_2, E_2}$ of the diagonal bars and the six random 
loads $\acc{P_1,\cdots,P_6}$. They are considered mutually independent and 
their distributions are given in Table~\ref{tbl:Truss input}. An in-house 
developed Matlab-based finite-element solver is used to calculate the 
displacement at midspan $u(\vX)$, counted positively downwards.

\begin{table}
	\caption{Two-dimensional truss structure: definition of the probabilistic 
		model of the input variables \citep{SchoebiASCE2016}}
	\label{tbl:Truss input}
	\small\centering
	\begin{tabular*}{.9\textwidth}{c@{\extracolsep{\fill}}c@{\extracolsep{\fill}}c@{\extracolsep{\fill}}c@{\extracolsep{\fill}}}
		
		\hline
		Variable &
		Distribution &
		Mean &
		Standard Deviation\\
		\hline
		$E_1$, $E_2$ (Pa)&
		Lognormal&
		$2.1 \times 10^{11}$&
		$2.1 \times 10^{10}$\\
		
		$A_1$ (m\textsuperscript{2})&
		Lognormal&
		$2.0 \times 10^{-3}$&
		$2.0 \times 10^{-4}$\\
		
		$A_2$ (m\textsuperscript{2})&
		Lognormal&
		$1.0 \times 10^{-3}$&
		$1.0 \times 10^{-4}$\\
		
		$P_1,\cdots,P_6$ (N)&
		Gumbel&
		$5.0 \times 10^{4}$&
		$7.5 \times 10^{3}$\\
		\hline
		
	\end{tabular*}
\end{table}

This structure operates in the nominal range as long as the midspan 
displacement is smaller than a critical threshold $\tau = 12$ cm, which 
can be cast as the following limit-state function:
\begin{equation}
\label{eqn:limit state function truss}
	g(\vx) = \tau - u(\vx)
\end{equation}
where $g(\vx) \leq 0$  if the system is in a failure state.

Because the FEM computational model is relatively cheap to evaluate, we could 
run a direct MCS-analysis with $N = 10^6$ samples to provide the reference $P_F 
= 1.52\cdot 10^{-3}$ for validation purposes. Additionally, standard FORM and 
SORM analyses were run to estimate the non-linearity of the limit-state 
surface. FORM underestimated the failure probability of a factor of almost 2 
and a cost of $N_{FORM} = 160$ model runs, while SORM achieved a good accuracy 
at a cost of $N_{SORM} = 372$ model runs, which suggests that the underlying 
problem is non-linear. Neither of the two methods, however, provides confidence 
interval on their estimates.

The A-bPCE algorithm was initialized with an experimental design consisting 
in a uniform sampling of a ball (for details, see \textit{e.g.} 
\citet{DubourgThesis}) of size $N_{ini} = 30$, while the sparse adaptive PCE was 
given a polynomial degree range $1 \leq p \leq 10$, hyperbolic truncation 
with q-norm $q = 0.75$ \citep{BlatmanJCP2011} and maximum allowed interaction 
$r = 2$ \citep{UQdoc_10_104}. The internal MCS sample size was 
$N_{MCS} = 10^6$ and the algorithm was set to add $K = 3$ new samples per 
iteration. The stopping criterion in Eq.~\eqref{eqn:Pf convergence} was set to 
$\epsilon_{\widehat{P}_F} = 0.10$.
For comparison purposes, we also ran a standard AK-MCS analysis with the same 
initial experimental design and convergence criterion. The covariance family of 
choice for the underlying Kriging model was chosen as Gaussian.

\begin{table}\small
	\centering
	\caption{Comparison of the estimation of $P_F$ with several algorithms for 
		the truss structure example. }
	\label{tbl:Results:Truss}
	
	\begin{tabular}{cccccc}
		\hline
		Algorithm &
		$\widehat{P}_F$ &
		$[\widehat{P}_F^-, \widehat{P}_F^+]$&
		$\widehat{\beta}$&
		$[\widehat{\beta}^-, \widehat{\beta}^+]$&
		$N_{\text{tot}}$\\
		\hline
		MCS (ref.) &
		$1.52\times 10^{-3}$&
		- &
		$2.96$&
		-&
		$10^6$\\
		
		FORM &
		$0.76\times 10^{-3}$&
		-&
		$3.17$&
		-&
		$160$\\
		
		SORM &
		$1.63\times 10^{-3}$&
		-&
		$2.94$&
		-&
		$372$\\
		
		AK-MCS &
		$1.52\times 10^{-3}$&
		$[1.44, 1.59]\times 10^{-3}$&
		$2.96$&
		$[2.90, 3.01]$&
		$300$\\
		A-bPCE &
		$1.48\times 10^{-3}$&
		$[1.43, 1.54]\times 10^{-3}$ &
		$2.97$&
		$[2.96, 2.98]$&
		$129$\\
		\hline
	\end{tabular}
	
\end{table}

Table~\ref{tbl:Results:Truss} presents a comparison of the estimated 
$\widehat{P}_F$ with the aforementioned analyses. Both AK-MCS and A-bPCE 
estimates of $P_F$ include the reference value within the confidence bounds set 
by the convergence criterion. However, for this particular example and choice 
of convergence criterion, A-bPCE achieved convergence significantly faster 
than AK-MCS, with a total cost of $129$ model evaluations, as compared to the 
$300$ required by AK-MCS, resulting in a final PCE of degree $p=3$ with $P = 
43$ basis elements. 

Overall, A-bPCE provides a stable estimate of the failure probability and 
confidence intervals at a cost that is lower than FORM for this example.

\subsection{Top-floor displacement of a structural frame}
\label{sec:Results:Frame}
\begin{figure}
	\centering
	\includegraphics[width=0.8\linewidth]{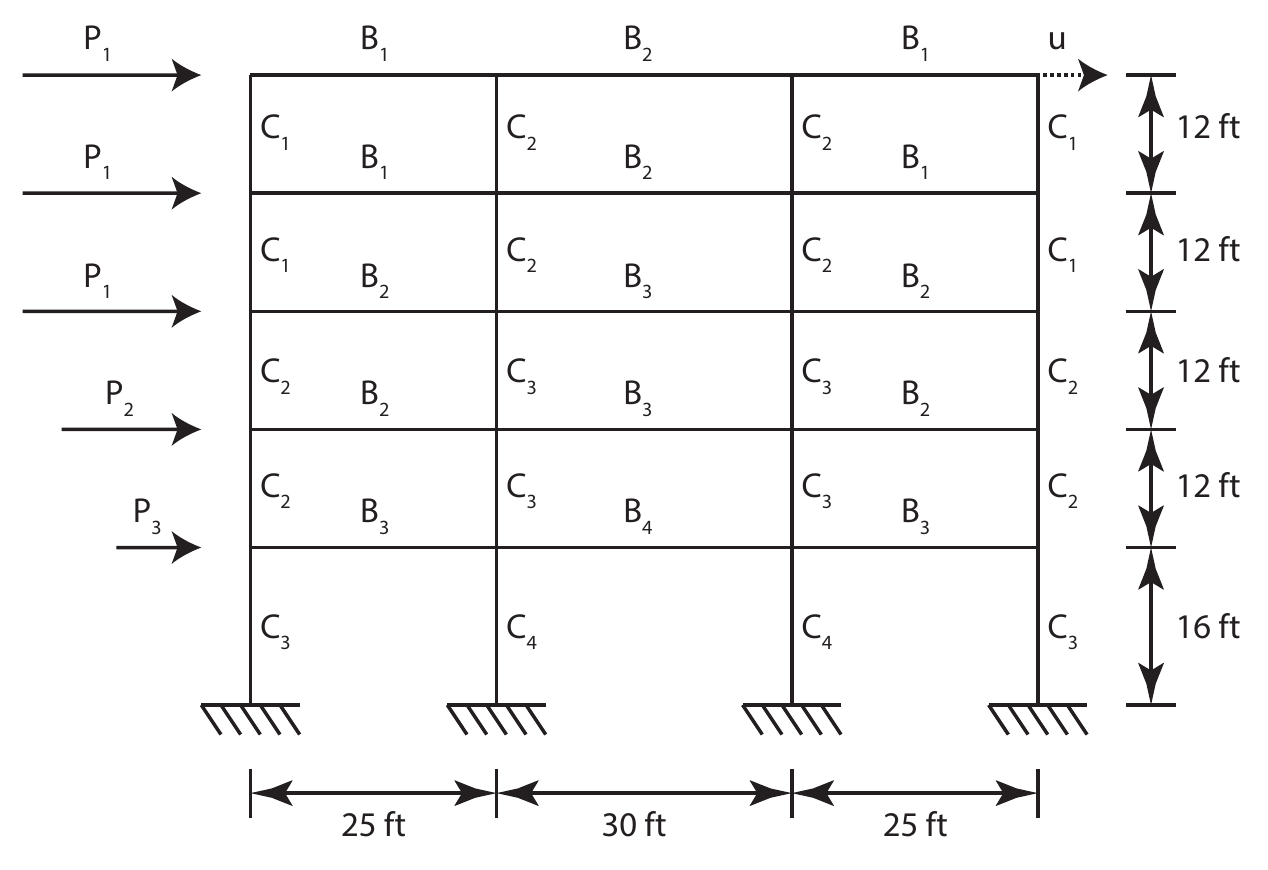}
	\caption{20-dimensional structural frame in 
	Section~\ref{sec:Results:Frame}. 
		The distributions of the input variables are reported in 
		Table~\ref{tbl:Results:FrameInputs}}.
	\label{fig:FrameStructure}
\end{figure}

Figure~\ref{fig:FrameStructure} shows a well known, high dimensional benchmark 
in structural reliability applications \citep{LiuPL91,BlatmanPEM2010}. It 
consists on a three-span, five story frame structure that is subject to 
horizontal loads. Both the loads and the properties of the elements of the 
frame (see Table~\ref{tbl:Results:FrameLabels}) are uncertain. Of interest is 
the top-floor horizontal displacement at the top right corner $u$.

\begin{table}
  \caption{Frame structure: properties of the elements shown in 
    Figure~\ref{fig:FrameStructure}}
  \label{tbl:Results:FrameLabels} \centering
	\begin{tabular*}{.8\textwidth}{c@{\extracolsep{\fill}}c@{\extracolsep{\fill}}c@{\extracolsep{\fill}}c@{\extracolsep{\fill}}}
          \hline
           Element &
           Young's modulus&
           Moment of inertia &
          Cross-sectional area\\\hline
		
          $B_1$ & $E_4$ & $I_{10}$ & $A_{18}$\\
          $B_2$ & $E_4$ & $I_{11}$ & $A_{19}$\\
          $B_3$ & $E_4$ & $I_{12}$ & $A_{20}$\\	
          $B_4$ & $E_4$ & $I_{13}$ & $A_{21}$\\
		
          $C_1$ & $E_5$ & $I_{6}$ & $A_{14}$\\
          $C_2$ & $E_5$ & $I_{7}$ & $A_{15}$\\
          $C_3$ & $E_5$ & $I_{8}$ & $A_{16}$\\	
          $C_4$ & $E_5$ & $I_{9}$ & $A_{17}$\\
		
		\hline
	\end{tabular*}
	
\end{table}

The uncertainties on the applied loads $P_1,...,P_3$, the Young's moduli 
$E_1$ and $E_2$, the moments of inertia $I_6,...,I_{13}$ and the cross sections 
$A_{14},...,A_{21}$ are modelled by a 21-dimensional joint random vector 
$\ve{Z} = \acc{P_1,...,A_{21}}$ with marginal distributions given in 
Table~\ref{tbl:Frameinput}. 

\begin{table}
	\caption{Frame structure: definition of the probabilistic 
		model of the input variables \citep{BlatmanPEM2010}}\label{tbl:Frameinput}
	\label{tbl:Results:FrameInputs}
	\small
	\begin{center}
		\begin{tabular*}{.9\textwidth}{c@{\extracolsep{\fill}}c@{\extracolsep{\fill}}c@{\extracolsep{\fill}}c@{\extracolsep{\fill}}}
			\hline
			Variable &
			Distribution &
			Mean &
			Standard Deviation\\
			\hline
			
			$P_1$ (kN)&
			Lognormal&
			$133.454$&
			$40.04$\\
			
			$P_2$ (kN)&
			Lognormal&
			$88.97$&
			$35.59$\\
			
			$P_3$ (kN)&
			Lognormal&
			$71.175$&
			$28.47$\\
			
			$E_4$ (kN/m\textsuperscript{2})&
			Truncated Gaussian\textsuperscript{*} &
			$2.1738\times10^7$&
			$1.9152\times10^6$\\
			
			$E_5$ (kN/m\textsuperscript{2})&
			Truncated Gaussian\textsuperscript{*} &
			$2.3796\times10^7$&
			$1.9152\times10^6$\\
			
			$I_6$ (m\textsuperscript{4})&
			Truncated Gaussian\textsuperscript{*} &
			$8.1344\times10^{-3}$&
			$1.0834\times10^{-3}$\\
			
			$I_7$ (m\textsuperscript{4})&
			Truncated Gaussian\textsuperscript{*}&
			$1.1509\times10^{-2}$&
			$1.2980\times10^{-3}$\\
			
			$I_8$ (m\textsuperscript{4})&
			Truncated Gaussian\textsuperscript{*}&
			$2.1375\times10^{-2}$&
			$2.5961\times10^{-3}$\\
			
			$I_9$ (m\textsuperscript{4})&
			Truncated Gaussian\textsuperscript{*}&
			$2.5961\times10^{-2}$&
			$3.0288\times10^{-3}$\\
			
			$I_{10}$ (m\textsuperscript{4})&
			Truncated Gaussian\textsuperscript{*}&
			$1.0812\times10^{-2}$&
			$2.5961\times10^{-3}$\\
			
			$I_{11}$ (m\textsuperscript{4})&
			Truncated Gaussian\textsuperscript{*}&
			$1.4105\times10^{-2}$&
			$3.4615\times10^{-3}$\\
			
			$I_{12}$ (m\textsuperscript{4})&
			Truncated Gaussian\textsuperscript{*}&
			$2.3279\times10^{-2}$&
			$5.6249\times10^{-3}$\\
			
			$I_{13}$ (m\textsuperscript{4})&
			Truncated Gaussian\textsuperscript{*}&
			$2.5961\times10^{-2}$&
			$6.4902\times10^{-3}$\\
			
			$A_{14}$ (m\textsuperscript{2})&
			Truncated Gaussian\textsuperscript{*}&
			$3.1256\times10^{-1}$&
			$5.5815\times10^{-2}$\\
			
			$A_{15}$ (m\textsuperscript{2})&
			Truncated Gaussian\textsuperscript{*}&
			$3.7210\times10^{-1}$&
			$7.4420\times10^{-2}$\\
			
			$A_{16}$ (m\textsuperscript{2})&
			Truncated Gaussian\textsuperscript{*}&
			$5.0606\times10^{-1}$&
			$9.3025\times10^{-2}$\\
			
			$A_{17}$ (m\textsuperscript{2})&
			Truncated Gaussian\textsuperscript{*}&
			$5.5815\times10^{-1}$&
			$1.1163\times10^{-1}$\\
			
			$A_{18}$ (m\textsuperscript{2})&
			Truncated Gaussian\textsuperscript{*}&
			$2.5302\times10^{-1}$&
			$9.3025\times10^{-2}$\\
			
			$A_{19}$ (m\textsuperscript{2})&
			Truncated Gaussian\textsuperscript{*}&
			$2.9117\times10^{-1}$&
			$1.0232\times10^{-1}$\\
			
			$A_{20}$ (m\textsuperscript{2})&
			Truncated Gaussian\textsuperscript{*}&
			$3.7303\times10^{-1}$&
			$1.2093\times10^{-1}$\\
			
			$A_{21}$ (m\textsuperscript{2})&
			Truncated Gaussian\textsuperscript{*}&
			$4.1860\times10^{-1}$&
			$1.9537\times10^{-1}$\\
			
			\hline
			
		\end{tabular*}
	\end{center}
	\begin{itemize}
		\item[\textsuperscript{*}] Truncated in the domain $[0, +\infty]$. The 
		quoted moments refer to the full, untruncated Gaussian distributions.
	\end{itemize}
\end{table}

Additionally, a Gaussian copula 
\citep{Lebrun2009c} is used to model dependence between the variables. The 
elements of the Gaussian copula correlation matrix $\ve{R}$ are given as:
\begin{itemize}
	\item $R_{E_1,E_2} = 0.9$ -- the two Young's moduli are highly correlated;
	\item $R_{A_i,I_i} = 0.95$ -- each element's cross-sectional area is highly 
	correlated to the corresponding moment of inertia;
	\item ${R_{A_i,I_j} = R_{I_i,I_j} = R_{A_i,A_j} = 0.13}$ -- the correlation 
	between the properties of different elements is much 
	lower;
	\item All the remaining elements of $\ve{R}$ are set to $0$.
\end{itemize}

A critical displacement of $\tau = 5$cm is identified as the maximum admissible 
threshold for the displacement $u$, hence resulting in the limit-state function:
\begin{equation}
\label{eqn:limit state function frame}
g(\ve{z}) = \tau - u(\ve{z})
\end{equation}
where $u(\ve{z})$ is the displacement on the top right corner calculated with 
an in-house FEM code. Due to the associated computational costs, the maximum 
available budget for the calculation of a reference solution is in this case 
limited to $N_{REF} = 4\cdot10^4$. Therefore, the reference solution is 
calculated with standard importance sampling (IS) \citep{Melchers1999} instead 
of direct MCS. In addition to Importance sampling, we also ran FORM and SORM. 
Due to the non-linearity of the problem, FORM significantly underestimated 
$P_F$, while SORM provided an accurate estimate. However, due to the high 
dimensionality of the input space, the associated cost in terms of model 
evaluation was relatively high, with $N_{SORM} = 1146$ model runs, since all 
the gradients of the limit-state function are computed using finite-differences.

The A-bPCE algorithm was initialized with an experimental design consisting 
of an LHS sampling of the input random vector of size $N_{ini} = 40$. Sparse 
PCE was carried out with a q-norm truncation with $q = 0.75$  and maximum 
allowed interaction $r = 2$. Note that the initialization is essentially the 
same as for the truss structure in the previous application. The internal MCS 
sample size was $N_{MCS} = 10^6$, with single point enrichment per iteration. 
The stopping criterion in Eq.~\eqref{eqn:Pf convergence} was in this case set 
to $\epsilon_{\widehat{P}_F} = 0.15 $.
For comparison purposes, an AK-MCS analysis was also run on the same initial 
design, with similar settings and a Gaussian covariance family. 

A comparison of the results is gathered in Table~\ref{tbl:Results:Frame}. Due 
to the different estimation method between the reference probability 
(importance sampling) and the active learning-based methods (which rely on an 
inner MCS), no direct comparison of the results is possible as in the previous 
cases. Indeed, even fixing the same random seeds would result in different 
estimates due to the different methodologies. Therefore, confidence bounds are 
given for all the three methods: 95\% confidence bounds for 
IS \citep{Melchers1999}, and $P_F^{\pm}$ for both AK-MCS and A-bPCE. The three 
methods give comparable results, albeit with significant differences in the 
convergence behaviour. In particular, both AK- and A-bPCE resulted in a 
slight underestimation of the probability of failure w.r.t. the reference 
solution by IS, which in turn is slightly overestimated with respect to the 
reference result quoted in the literature \citep{BlatmanPEM2010}. However, AK-MCS 
did not converge in the allotted maximum number of model evaluations, and its 
confidence bounds remained remarkably large with respect to A-bPCE. A-bPCE 
converged instead at a total cost of approximately $200$ model evaluations to 
the target $\epsilon_{\widehat{P}_F}$, with a final sparse PCE of degree 2, 
counting $30$ non-zero coefficients. 
For both active-learning-based methods, the reference solution lies within the 
given confidence bounds.
Moreover, the confidence bounds on the reliability index $\widehat{\beta}$ show 
that the results are stable to within $2\%$ of the calculated values.

Finally, it is interesting to mention that for this example the costs of FORM 
and A-bPCE were comparable, but the latter provides a much less biased 
estimate, and includes confidence bounds.

\begin{table}
	\caption{Comparison of the estimation of $P_F$ with several algorithms for 
	the top-floor displacement of a structural frame example }
		\label{tbl:Results:Frame}
		\begin{center}
			\begin{tabular}{cccccc}
				\hline
				Algorithm&
				$\widehat{P}_F$&
				$[\widehat{P}_F^-, \widehat{P}_F^+]$&
				$\widehat{\beta}$&
				$[\widehat{\beta}^-, \widehat{\beta}^+]$&
				$N_{tot}$\\
				\hline
				IS (ref.) &
				$1.54\times 10^{-3}$&
				$[1.51 , 1.56]\times 10^{-3}$ &
				$2.96$&
				$[2.95, 2.97]$ &
				$40241$\\
				
				FORM &
				$1.01\times 10^{-3}$&
				-&
				$3.09$&
				$-$&
				$241$\\

				SORM &
				$1.52\times 10^{-3}$&
				-&
				$2.96$&
				$-$&
				$1146$\\

				AK-MCS &
				$1.48\times 10^{-3}$&
				$[0.8, 4.96]\times10^{-3}$ &
				$2.97$&
				$[2.57,3.12]$&
				$300$\\
				
				A-bPCE &
				$1.49\times 10^{-3}$&
				$[1.42, 1.62]\times10^{-3}$&
				$2.97$&
				$[2.94, 2.98]$&
				$235$
				\\
				\hline
			\end{tabular}
		\end{center}
\end{table}

\section{Conclusions and outlook}
\label{sec:Conclusions}

A novel approach to solving reliability problems with polynomial chaos 
expansions has been proposed. The combination of the bootstrap method and 
sparse regression enabled us to introduce local error estimation in the 
standard PCE predictor. 
In turn, this allows one to construct active learning algorithms similar to 
AK-MCS to greedily enrich a relatively small initial experimental design so as 
to efficiently estimate the probability of failure of complex systems. 

This approach has shown comparable performance \textit{w.r.t.} to the well 
established AK-MCS method on both a simple analytical benchmark function and in 
two high-dimensional engineering applications of increasing complexity.

Extensions of this approach can be envisioned in two main directions: 
\begin{itemize}
	\item the simulation-based reliability analysis method can be extended 
	beyond simple MCS (\textit{e.g.} by using importance sampling 
	\citep{Dubourg2013}, line sampling \citep{Pradlwarter2007} or subset 
	simulation 	\citep{Dubourg2011}) to achieve better $\widehat{P}_F$ estimates 
	at each iteration, especially for very low probabilities of failure;
	\item remote parallel computing facilities may be used during the 
	enrichment phase of the algorithms with expensive computational models when 
	adding more than one point at a time;
	\item the use of bootstrap to enable local error estimation in an active 
	learning context can be used also with different regression-based surrogate 
	modelling techniques, including \textit{e.g.} low rank tensor 
	approximations \citep{KonakliPEM2016}.
\end{itemize}

Additionally, the bPCE approach itself introduced in this work can be used also 
outside of a pure reliability analysis context, as it provides an effective 
local error estimate for PCE. It has been used, \textit{e.g.} in the context of 
reliability-based design optimization in \citet{MoustaphaICOSSAR2017}. 
Indeed the lack of this feature (as opposed to Kriging) has somewhat hindered 
its usage in more advanced active-learning applications.

\section*{References}

\bibliography{MarelliSudret_bPCEStructSafety_R1}

\end{document}